\documentclass[10pt,]{IEEEtran}
\usepackage{amssymb}
\usepackage{color}
\usepackage{graphicx}
\usepackage{amsmath}
\usepackage{wasysym}
\usepackage{multirow}
\usepackage{booktabs}
\usepackage{lineno}
\usepackage{soul}
\usepackage[draft]{fixme}
\usepackage{longtable}
\usepackage{supertabular}
\usepackage{mathtools}
\usepackage{caption}
\usepackage{subcaption}
\usepackage{authblk}
\usepackage[colon]{natbib}
\newcommand{\delOx}{$\delta{}^{18}\mathrm{O}$ }
\newcommand{\delD}{$\delta\mathrm{D}$ }

\begin{document}

%\begin{frontmatter}

%% Title, authors and addresses

%% use the tnoteref command within \title for footnotes;
%% use the tnotetext command for theassociated footnote;
%% use the fnref command within \author or \address for footnotes;
%% use the fntext command for theassociated footnote;
%% use the corref command within \author for corresponding author footnotes;
%% use the cortext command for theassociated footnote;
%% use the ead command for the email address,
%% and the form \ead[url] for the home page:
%% \title{Title\tnoteref{label1}}
%% \tnotetext[label1]{}
%% \author{Name\corref{cor1}\fnref{label2}}
%% \ead{email address}
%% \ead[url]{home page}
%% \fntext[label2]{}
%% \cortext[cor1]{}
%% \address{Address\fnref{label3}}
%% \fntext[label3]{}

\title{Supporting Online Material to the manuscript: ``Water isotope diffusion rates from the NorthGRIP ice core for the last 16,000
years - glaciological and paleoclimatic implications.''}

%% use optional labels to link authors explicitly to addresses:
%% \author[label1,label2]{}
%% \address[label1]{}
%% \address[label2]{}

\author[1,2]{V. Gkinis}
\author[1,3]{S. B. Simonsen}
\author[1]{S. L. Buchardt}
\author[2]{J. W. C. White}
\author[1]{B. M. Vinther}

%\address[1]{Centre for Ice and Climate, Niels Bohr Institute, University of Copenhagen,
%Juliane Maries Vej 30, DK-2100 Copenhagen, Denmark}
%\address[2]{Institute for Alpine and Arctic Research, University of Colorado, Boulder, 1560 30th Street
%Boulder, CO 80303 USA}
%\address[3]{Div. of Geodynamics, DTU space � National Space Institute,
%Elektrovej, Build. 327, Kgs. Lyngby, Denmark}
\affil[1]{Centre for Ice and Climate, Niels Bohr Institute, University of Copenhagen,
Juliane Maries Vej 30, DK-2100 Copenhagen, Denmark}
\affil[2]{Institute for Alpine and Arctic Research, University of Colorado, Boulder, 1560 30th Street
Boulder, CO 80303 USA}
\affil[3]{Div. of Geodynamics, DTU space � National Space Institute,
Elektrovej, Build. 327, Kgs. Lyngby, Denmark}
\maketitle

%\begin{abstract}
%
%   	\noindent A high resolution (0.05 m) water isotopic record (\delOx) is available
%   	from the NorthGRIP ice core. In this study we look into the water isotope
%   	diffusion history as estimated by the spectral characteristics of the \delOx
%   	time series covering the last 16,000 years. Based on it we infer a temperature
%	history signal for the site.
%
%\end{abstract}

%\end{frontmatter}

%%%%%%%%%%%%%%%%%%%%%%%%%%%%%%%%%%%%%%%%%%%%%%%%%%
%%%%%%%%%%%%%%%%%%%%%%%%%%%%%%%%%%%%%%%%%%%%%%%%%%
%\tableofcontents

%%%%%%%%%%%%%%%%%%%%%%%%%%%%%%%%%%%%%%%%%%%%%%%%%%
%%%%%%%%%%%%%%%%%%%%%%%%%%%%%%%%%%%%%%%%%%%%%%%%%%
%\tableofcontents
%\noindent \rule{\linewidth}{0.4pt}
\section{Derivation of the integration equations for $\sigma^2$}

	In this section we show how we derive the numerical expressions for the diffusion length (Eq. (8) in main text (MT)).
	Starting from the differential equation for the diffusion length we have:
\begin{equation}
\frac{\mathrm{d}\sigma^2}{\mathrm{d}t} - 2\,\dot{\varepsilon}_z\!\left( t \right) \sigma^2 =
2D\!\left( t \right).
\label{diff_len_1}
\end{equation}
	We consider the vertical strain rate due to densification:
\begin{equation}
\dot{\varepsilon}_z \left(
	t \right) = \frac{-\partial \rho}{\,\,\,\partial t}\,\frac{\,1\,}{\,\rho\,}
\label{diff_len_2}
\end{equation}
	By substituting  $t$ with $\rho$ and combining Eq.(\ref{diff_len_1}), (\ref{diff_len_2}) we get:
\begin{equation}
\frac{\mathrm{d}\sigma^2}{\mathrm{d}\rho} + \frac{2\sigma^2}{\rho} = 2 {\left(\frac{\text{d}\rho}{\text{d}t}\right)}^{-1} D(\rho)
\label{diff_len_3}
\end{equation}
	Multiplication of both sides of Eq.\ref{diff_len_3} with the integrating factor
\begin{equation}
	F(\rho) = e^{\int \frac{2}{\rho}\text{d}\rho} = \rho^2,
\label{diff_len_4}
\end{equation}
	gives:
\begin{equation}
	\frac{\text{d}}{\text{d}t} \left( \rho^2\sigma^2 \right) = 2\rho^2 {\left( \frac{\text{d}\rho}{\text{d}t} \right)}^{-1} D(\rho),
\label{diff_len_5}
\end{equation}
	from which we get the result:
\begin{equation}
\sigma^2 \left( \rho \right) = \frac{\,1\,}{\rho^2}\int_{\rho_o}^{\rho}2\rho^2
{\left( \frac{\mathrm{d}\rho}{\mathrm{d}t}\right)}^{-1}\! D \!\left( \rho \right) \,\mathrm{d}\rho.
\label{diff_len_6}
\end{equation}

	In a similar way for the ice diffusion length (Eq. (12) in the MT) we have:
\begin{equation}
\frac{\mathrm{d}\sigma^2}{\mathrm{d}t} - 2\,\dot{\varepsilon}_z\!\left( t \right) \sigma^2 =
2D\!\left( t \right),
\label{diff_len_7}
\end{equation}
	where the total thinning is given by:
\begin{equation}
\mathcal{S} \left( t' \right) = e^{\int_0^{t'} \dot{\varepsilon}_z \left( t \right) \mathrm{d}t} \enspace .
\label{diff_len_8}
\end{equation}
	We multiply  both sides of Eq. (\ref{diff_len_7}) with the integrating factor
\begin{equation}
	F(\rho) = e^{\int_0^{t'}-2\dot{\varepsilon}_z \left( t \right) \mathrm{d}t},
\label{diff_len_9}
\end{equation}
	which results in
\begin{equation}
	\frac{\text{d}}{\text{d}t} \left[\sigma^2   e^{\int_0^{t'}-2\dot{\varepsilon}_z \left( t \right) \mathrm{d}t} \right]
	= 2D(t)e^{\int_0^{t'}-2\dot{\varepsilon}_z \left( t \right) \mathrm{d}t},
\label{diff_len_10}
\end{equation}
	From this, we get the expression for the ice diffusion length
\begin{equation}
\sigma_{ice}^2 \!\left( t' \right) =
S\! \left( t' \right)^2 \,\int_0^{t'} 2 D_{ice} \!\left( t \right) S \!\left( t \right) ^{-2} \,\mathrm{d} t.
\label{diff_len_11}
\end{equation}

%%%%%%%%%%%%%%%%%%%%%%%%%%%%%%%%%%%%%%%%%%%%%%%%%%
%%%%%%%%%%%%%%%%%%%%%%%%%%%%%%%%%%%%%%%%%%%%%%%%%%

\section{The diffusivity parametrization} \label{D_appendix}
\subsection{The firn diffusivity}
We use the diffusivity parametrization as introduced by \cite{Johnsen2000}.
\begin{equation}
D \!\left(\rho\right) = \frac{m\,p\,D_{ai}}{R\,T\,\alpha_i\,\tau}
\left(\frac{\,1\,}{\,\rho\,} - \frac{\,1\,}{\,\rho_{ice}\,}\right).
\label{eq2.8}
\end{equation}
	The terms used in Eq. (\ref{eq2.8}) and their parameterizations used are described below:
\begin{itemize}

\item{$m$: molar weight (kg)}

\item{$p$: saturation vapor pressure over ice (Pa). We use \citep{Murphy2005}:
\begin{equation}
p = \exp \left(9.5504 - \frac{5723.265}{T} + 3.530\,\ln\!\left( T \right) - 0.0073\,T \right).
\label{eq2.9}
\end{equation}}

\item{$D_{a}$:  diffusivity of water vapor in air ($\mathrm{m}^2 \mathrm{s}^{-1}$). We use \citep{Hall1976}:
\begin{equation}
D_{a} = 2.1\cdot 10^{-5} {\left(\frac{T}{T_o}\right)}^{1.94} \left(\frac{P_o}{P}\right)
\label{eq2.10}
\end{equation}
with $P_o = 1$ Atm, $T_o = 273.15$ K and $P, \;T$ the ambient pressure (Atm) and temperature (K).
Additionally from \cite{Merlivat1978} $D_{a^2\mathrm{H}} = \frac{D_a}{1.0251}$ and
$D_{a^{18}\mathrm{O}} = \frac{D_a}{1.0285}$}.

\item{$R$: molar gas constant $R = 8.314478 \,\,\left(\mathrm{m}^3\mathrm{Pa}\,\left(\mathrm{K \,mol}\right)^{-1}\right)$}

\item{$T$: Ambient temperature (K)}

\item{$\alpha_i$: Ice -- Vapor fractionation factor.  we use the formulations by \cite{Majoube1971}
and \cite{Merlivat1967} for $\alpha_{s/v}^{2}$ and $\alpha_{s/v}^{18}$ respectively.
\begin{align}
\ln \alpha_{Ice/Vapor} \left(^{2}\mathrm{H}/^{1} \mathrm{H} \right) &= 16288/T^2 - 9.34 \times 10^{-2}\\
\ln \alpha_{Ice/Vapor} \left( ^{18}\mathrm{O}/^{16} \mathrm{O} \right) &= 11.839/T - 28.224 \times 10^{-3}
\label{eq2.12}
\end{align}}

\item{$\tau$: The firn tortuosity. We use \citep{Schwander1988, Johnsen2000}:
\begin{equation}\label{eq2.13}
\frac{1}{\tau } =
\left\{
  \begin {array}{ll}
1 - b_\tau  \left( {\frac{\rho }{{\rho _{ice} }}} \right)^2 & \mbox{, for } \rho  \le \frac{\rho_{ice}}{\sqrt{b}}  \\
0 & \mbox{, for } \rho  > \frac{\rho_{ice}}{\sqrt{b}}\\
 \end{array}\right. \enspace,
\end{equation}
	where $b_\tau = 1.30$, implying a close-off density of $\rho_{\mathrm{co}} = 804.3\;\mathrm{kgm}^{-3}$.}

\end{itemize}

\subsection{The ice diffusivity}
\begin{figure*}
\vspace*{1mm}
\center
\includegraphics[width = 100mm]{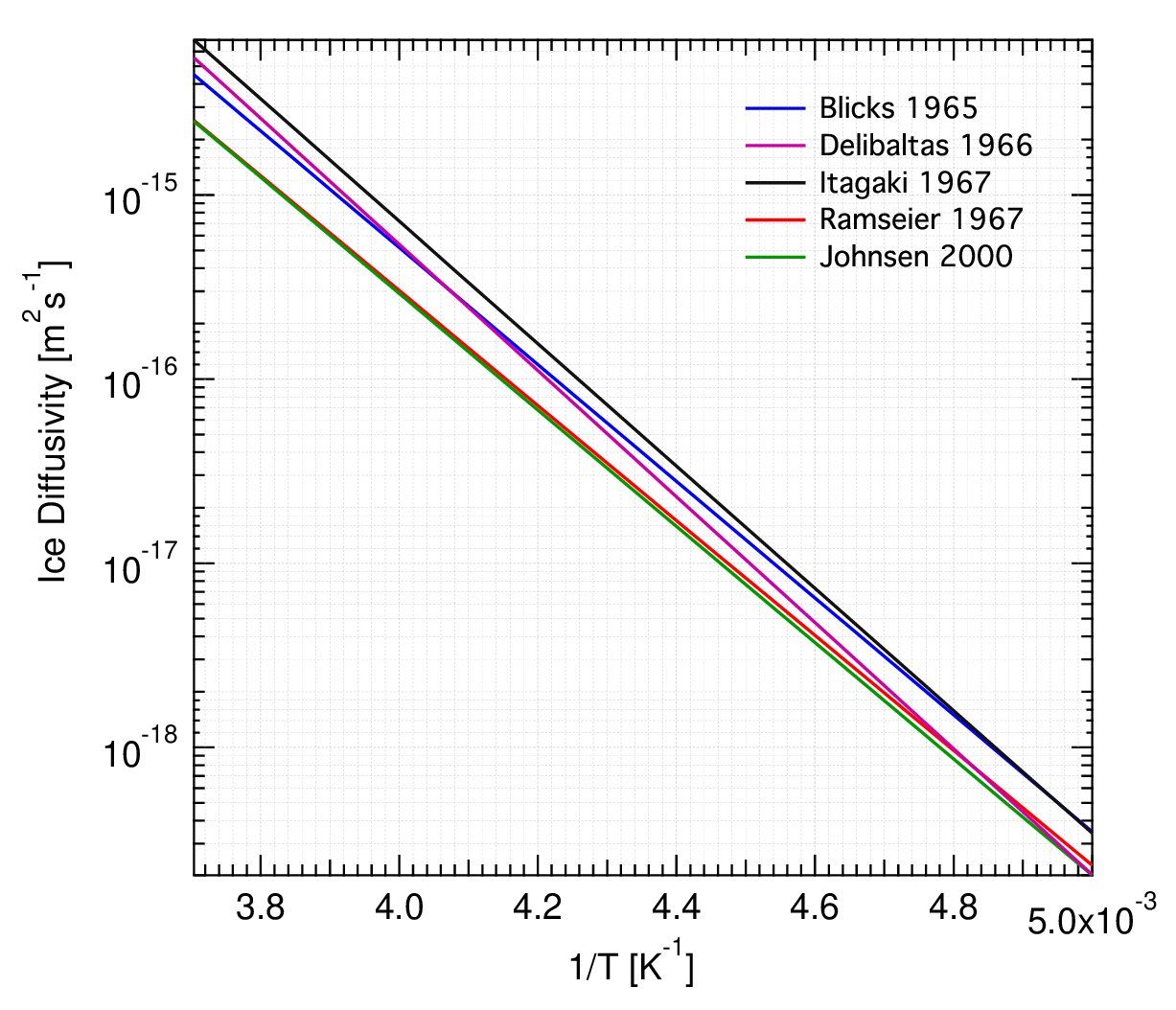}
\caption{Ice diffusivity parametrizations based on isotopic probe experiments
for the temperature range 200 -- 270 K}
\label{ice_diffusivities_plot}
\end{figure*}

	Ice diffusion  is believed to occur via a vacancy mechanism
	with transport of molecules within the ice lattice.
	Based on isotopic probe experiments, there is a strong consensus that
	the ice diffusivity coefficient is  the same for
	$\mathrm{H}_2 \,^{18}\mathrm{O}, \, \mathrm{D}_2 \mathrm{O} \text{ and } \mathrm{T}_2 \mathrm{O}$
	\citep{Ramseier1967, Blicks1966, Itagaki1967, Delibaltas1966}
	The dependence of the ice diffusivity parameter to temperature is described by
	an Arrhenius type equation
\begin{equation}
D = D_0 \exp \left( -Q/RT \right),
\label{arrhenius}
\end{equation}
	where $Q$ is the activation energy and $D_0$ a pre-exponential factor.
	The results of the studies mentioned above agree well with each other.
	Here we
	plot the diffusivity parametrization coefficients suggested by those studies (Fig. \ref{ice_diffusivities_plot}).
	In this work we follow \cite{Ramseier1967}
	and  use $Q = 0.62 \; \mathrm{eV}$ and $D_{o} = 9.2\cdot10^{-4}  \; \mathrm{m^2\,s^{-1}}$.
	Note that the results of \cite{Ramseier1967} are based on measurements of both artificially as well
	as naturally grown ice   collected at Mendenhall glacier, Alaska.

	Enhanced ice diffusion rates have been proposed to be the cause of the excess diffusion
	observed in the Holocene section of the GRIP ice core \citep{Johnsen2000}.
	For the early Holocene
	part of the GRIP core, the authors of that study observed  higher diffusion rates than expected by
	the theory.
	In order to diminish the discrepancy between modeled and observed diffusion rates
	\cite{Johnsen2000} introduced  the term  ``excess ice diffusion'', referring to a possibly
	higher diffusivity coefficient due to isotopic exchange in the liquid phase on thin water films and ice
	crystal veins. However, the thickness of the water films and the diameter of the
	ice crystal veins required, are unrealistically high.

	Although the existence of an ``excess ice diffusion'' mechanism cannot be excluded based
	on the findings of our study, it should be mentioned that the diffusion model used in \cite{Johnsen2000}
	assumes an accumulation rate and temperature signal that is based on the
	\delOx record of the GRIP core. The GRIP \delOx signal is characterized by a rather flat curve throughout
	the Holocene showing no indication of an early Holocene optimum, a feature that is mostly
	due to ice sheet elevation effects ``masking'' the temperature change \citep{Vinther2009}.
	As a result, it is expected that the diffusion length calculations in \cite{Johnsen2000} would underestimate the
	diffusion signal throughout the Holocene.
	We conclude that the ``excess ice diffusion'' issue requires more work in the future.
	Considering that the ice diffusivity coefficient is very similar for all the isotopologues
	of water, a study focusing on the differential diffusion signal between \delOx and \delD
	would provide a better insight in the problem. An effort will be undertaken as soon as
	an adequately long Holocene section of the NEEM ice core is analyzed for both
	\delOx and \delD using dual isotope laser spectroscopy.

%%%%%%%%%%%%%%%%%%%%%%%%%%%%%%%%%%%%%%%%%%%%%%%%%%
%%%%%%%%%%%%%%%%%%%%%%%%%%%%%%%%%%%%%%%%%%%%%%%%%%
\section{Examples of diffusion lengths for different ice core sites}
\begin{figure*}
\vspace*{1mm}
\center
\includegraphics[width = 180mm]{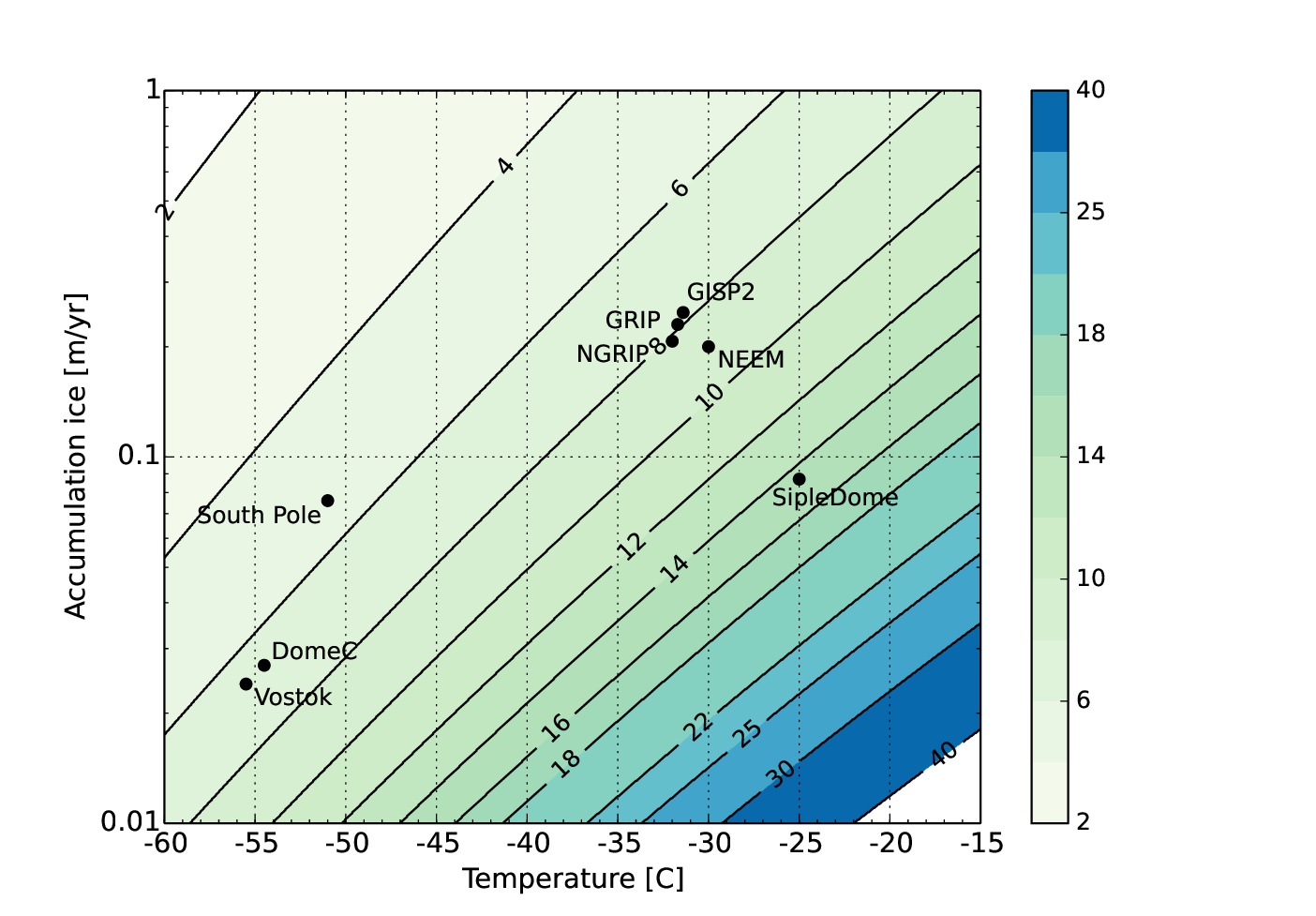}
\caption{Calculation of $\sigma_{18}^2$ for the close--off density of $\rho_{\text{co}} = 804.3\;\text{kgm}^{-3}$ in
m of ice equivalent. $\rho_o = 330\;\text{kgm}^{-3}$ for all sites.}
\label{diffusionlength_map}
\end{figure*}

	In this section we present an ensemble of implementations of the diffusion--densification model
	for various combinations of surface forcings that represent typical modern day conditions
	for a number of ice core sites on Greenland and Antarctica.
	The contours in the plot are generated by integration of Eq. (\ref{diff_len_6}) and expressed
	in m ice eq.
	The forcing for each ice core site is given in Table \ref{sigma_18_core_sites} and the
	results are shown in Fig. \ref{diffusionlength_map}.

%%%%%%%%%%%%%%%%%%%%%%%%%%%%%%%%%%%%%%%%%%%%%%%%%%
%%%%%%%%%%%%%%%%%%%%%%%%%%%%%%%%%%%%%%%%%%%%%%%%%%
\section{Estimation of $\sigma^2$ from the high resolution data set}
\label{sigma_estimation}

	In order to estimate the diffusion length value from high resolution water isotope data we minimize the
	2-norm $\| P_s - \hat{P}_s \|$ where $\hat{P}_s$ is an estimate of the power spectral density of
	a high resolution \delOx data section and $P_s$ is a model description of the power spectral density.

	$\hat{P}_s$ is obtained by the use
	of the Burg's spectral estimation method. The method fits an autoregressive model
	of order $\mu$ (AR-$\mu$) by minimizing the forward--backward prediction error filter
	\citep{Hayes1996, RECIPES, Andersen1974}.
	For the theoretical model  we have:
\begin{equation}
\label{powersd}
P_s =  P_{\sigma}  + {\vert \hat{\eta} \left( k \right) \vert} ^{2},
\end{equation}%e11
	where $P_{\sigma} = P_0 \,{e}^{-k^2 \sigma_i^2}$ is the effect of the
	firn diffusion process with squared diffusion length $\sigma^2$.
	Regarding the noise, we find red noise described by an AR-1 process with an autoregressive coefficient
	$q_1 = 0.15$ to provide a
	good description of the noise signal we observe. The spectrum of
	this signal is \citep{Kay1981}:
\begin{equation}
|\hat{\eta}(k )|^2 = \frac{\sigma_{\eta}^2 \Delta} {\left| 1+q_1 \exp{\left( - i k \Delta \right) } \right|^2} {},
\label{rednoise}
\end{equation}%e21
 	where $\sigma_{\eta}^2$ is the variance of the noise.
	The angular frequency $k=2\pi f$  is in the range $f \in \left[ 0, \; \frac{1}{2\Delta}  \right] $
	defined by the Nyquist frequency and thus the sampling resolution $\Delta$.
	We vary the parameters
	$\sigma^2, \, P_0, $ and $\sigma_{\eta}^2$ of the spectral model
	in order to minimize the misfit
	between $P_s$ and  $\hat{P}_s$ in a least squares sense.

\begin{figure*}
\vspace*{1mm}
\center
\includegraphics[width = 110mm]{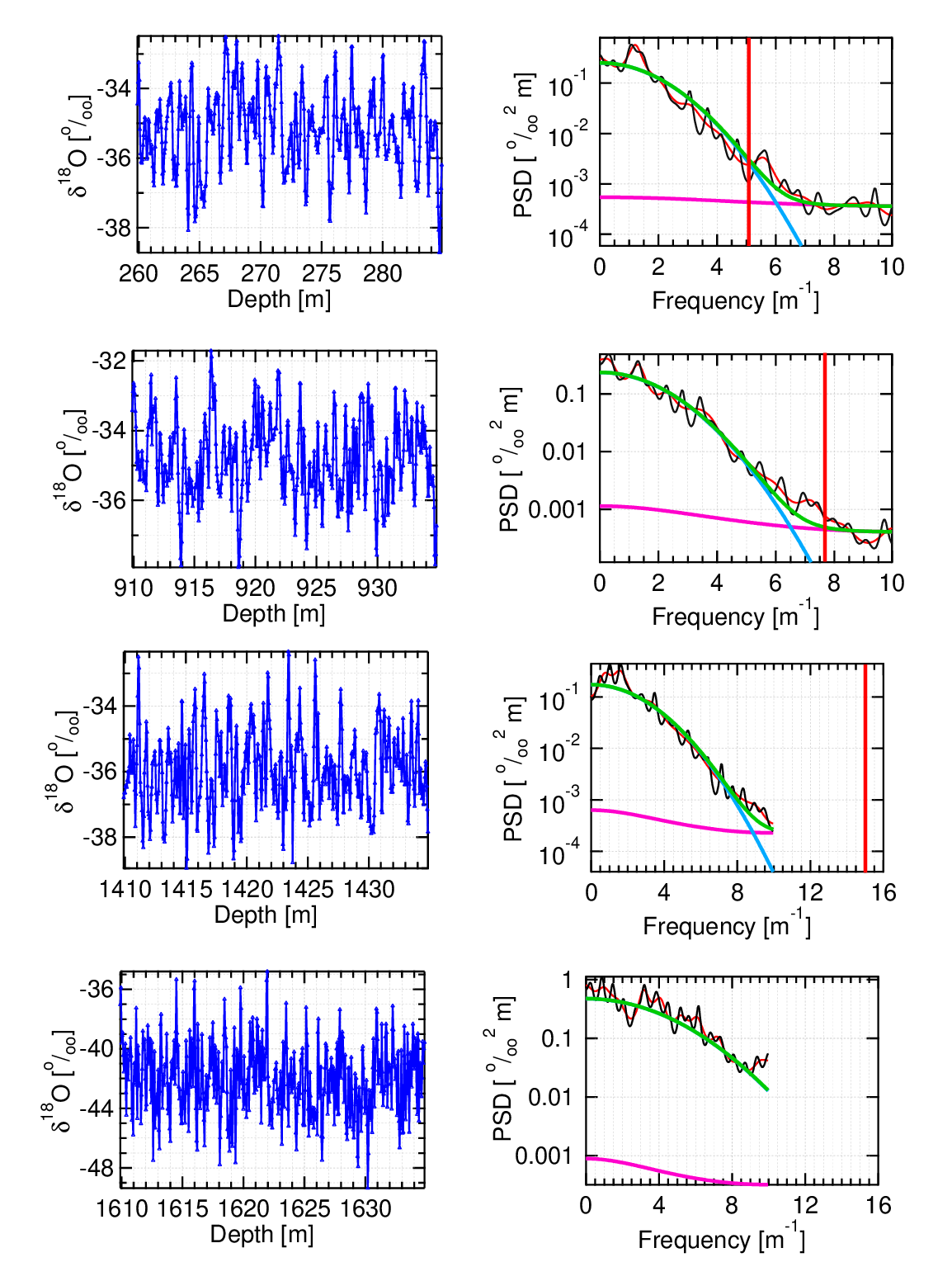}
\caption{Examples of raw \delOx data and estimated power spectral densities $\hat{P}_s$ for four  depth
intervals using $\mu = 30$ (red spectra) and $\mu = 40$ (black spectra).
The power spectral model is illustrated; $P_s$ in green, $P_{\sigma}$ in cyan and
${\vert \hat{\eta} \left( k \right) \vert} ^{2}$ in pink. The red vertical line in the top three plots
indicates the approximate position of the frequency representing the annual layer thickness.
For the bottom plot this frequency is $\approx 40 \, \mathrm{m}^{-1}$ and omitted from the plot.}
\label{spectral01}
\end{figure*}

	As it can be  seen in  Fig. \ref{spectral01}, it is the characterization of the full spectrum that yields
	information on $\sigma^2$. It is thus not necessary to specifically study the relative attenuation
	of individual spectral peaks as for example the annual signal. This approach allows for a study
	of the diffusion signal even after the spectral signature of the annual signal diminishes.

%%%%%%%%%%%%%%%%%%%%%%%%%%%%%%%%%%%%%%%%%%%%%%%%%%
%%%%%%%%%%%%%%%%%%%%%%%%%%%%%%%%%%%%%%%%%%%%%%%%%%
\section{AR order selection}
\label{AR_order_selection}
	An interesting feature of the Burg estimation method is that
	the order $\mu$ of the AR filter affects the spectral resolution of $\hat{P}_s$ \citep{Hayes1996, RECIPES}.
	Low $\mu$
	values result in smoother spectra with inferior spectral resolution, while higher order spectra show better
	performance in resolving neighboring spectral peaks. This can be seen in the spectral estimates
	presented in Fig.\ref{spectral01} where we plot spectral estimates with  $\mu = 30$ and
	$\mu = 40$.
	As described above, the goal of the $\sigma^2$ estimation is to characterize the overall shape of the spectrum.
	As a result, relatively low values of $\mu$
	produce smooth spectra of relatively low spectral resolution and can be
	adequate for the purpose of our application.

\begin{figure*}
\vspace*{1mm}
\center
\includegraphics[width = 100mm]{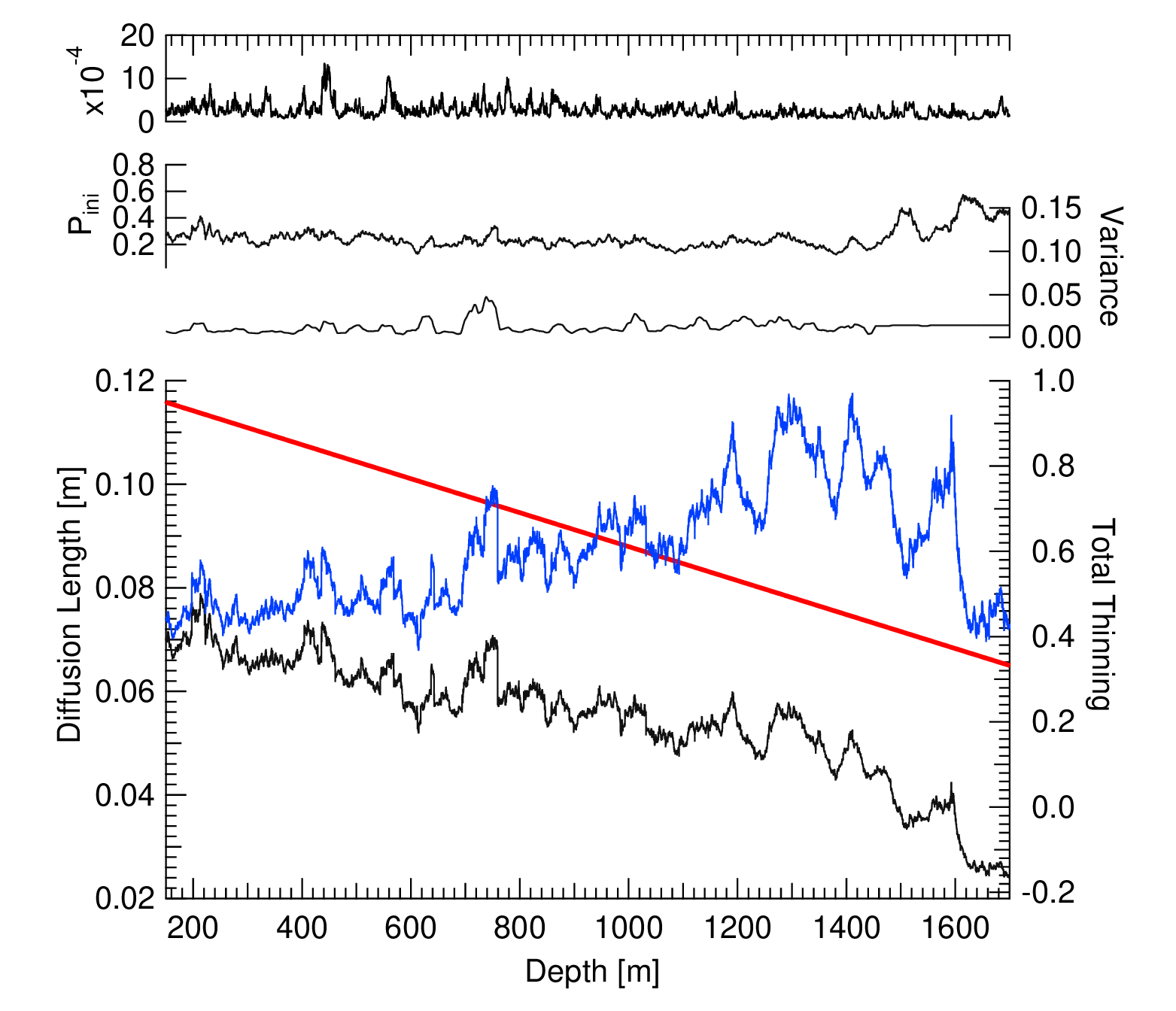}
\caption{Result of the AR order selection test (section \ref{AR_order_selection}).
	The black curve is the mean value of the diffusion length calculated using a $\mu$ in the
	range $\left[40, 80 \right]$. The blue curve is the diffusion length curve corrected for
	ice flow thinning (red curve) effects.
	The mean $P_0$ and ${\sigma_{\eta}}^2$ values are given in the two middle plots.
	In the top plot the standard deviation
	of the 41 estimates of the diffusion length for every depth in m ice eq. is shown.}
\label{spectral02}
\end{figure*}

	We look into both the influence of the value of $\mu$  on the $\sigma^2$ estimate
	by performing 41 power spectrum estimates with $\mu \in \left[ 40, 80 \right]$.
	Possible interferences of spectral features due to longer scale climate variability that could have
	an effect on the estimation of $\sigma^2$ are also investigated with this test.
	In Fig. \ref{spectral02} we show the mean value
	of the 41 spectral estimates before and after strain correction.
	The standard deviation of the estimated
	$\sigma$ for every depth is presented on the top subplot of the figure.
	It can be seen that on average
	the standard deviation is about 2 orders of magnitude lower than the absolute values of $\sqrt{\sigma^2}$
	thus  approximately in the 1\% range.
	The low standard deviation of the 41 estimates suggests that a possible
	effect of spectral features due to low frequency climate variability is of second order.
	The same applies for the selection of the AR order $\mu$, in Burg's spectral estimation.

%%%%%%%%%%%%%%%%%%%%%%%%%%%%%%%%%%%%%%%%%%%%%%%%%%
%%%%%%%%%%%%%%%%%%%%%%%%%%%%%%%%%%%%%%%%%%%%%%%%%%
\section{Test with synthetic data}
\label{synthetic_data}
	Additional to the MEM order selection sensitivity test shown in section \ref{AR_order_selection},
	we investigate the precision and accuracy  of the $\sigma^2$
	estimation using synthetic data.
	We perform two  tests which we describe below.
\subsection{Synthetic data test 1}
	For the first test we investigate the influence of short or long memory of the
	\delOx time series due to climate variability.
	We generate high resolution synthetic \delOx data by assuming an AR--1 process with the
	AR--1 coefficient $\phi_1$ of the process being equal to 0.2 and 0.9995.
	The process is applied on Gaussian noise with a mean of zero and a
	standard deviation of $10\, \permil$.
	The time series generated have a spacing of $\Delta x = 10^{-3} \,\mathrm{m}$ and a total
	length $L = 20 \, \mathrm{m}$.  A $\phi_1$ equal to 0.9995 with  $\Delta x = 10^{-3} \,\mathrm{m}$
	corresponds to a time constant for the memory of the process equal to
	$\tau_{0.9995} = -\Delta x / \ln \phi_1 = 2 \,\mathrm{m}$
	\citep{Percival1993}.
	For the part of the record we study, this is equivalent to climate
	variability at decadal time scales.
	The high resolution AR--1 process is then convolved
	with the Gaussian filter of predefined variance, simulating the effect of firn diffusion.
	Measurement white noise is then added to the result of the convolution.
	We then sample the high resolution diffused time series at a  resolution of
	$\Delta = 0.05 \,\mathrm{m}$ representing the width of a discrete  ice core sample
	and perform the $\sigma^2$ estimation as
	described in section \ref{sigma_estimation}.

\begin{figure*}
\vspace*{1mm}
\center
\includegraphics[width = 150mm]{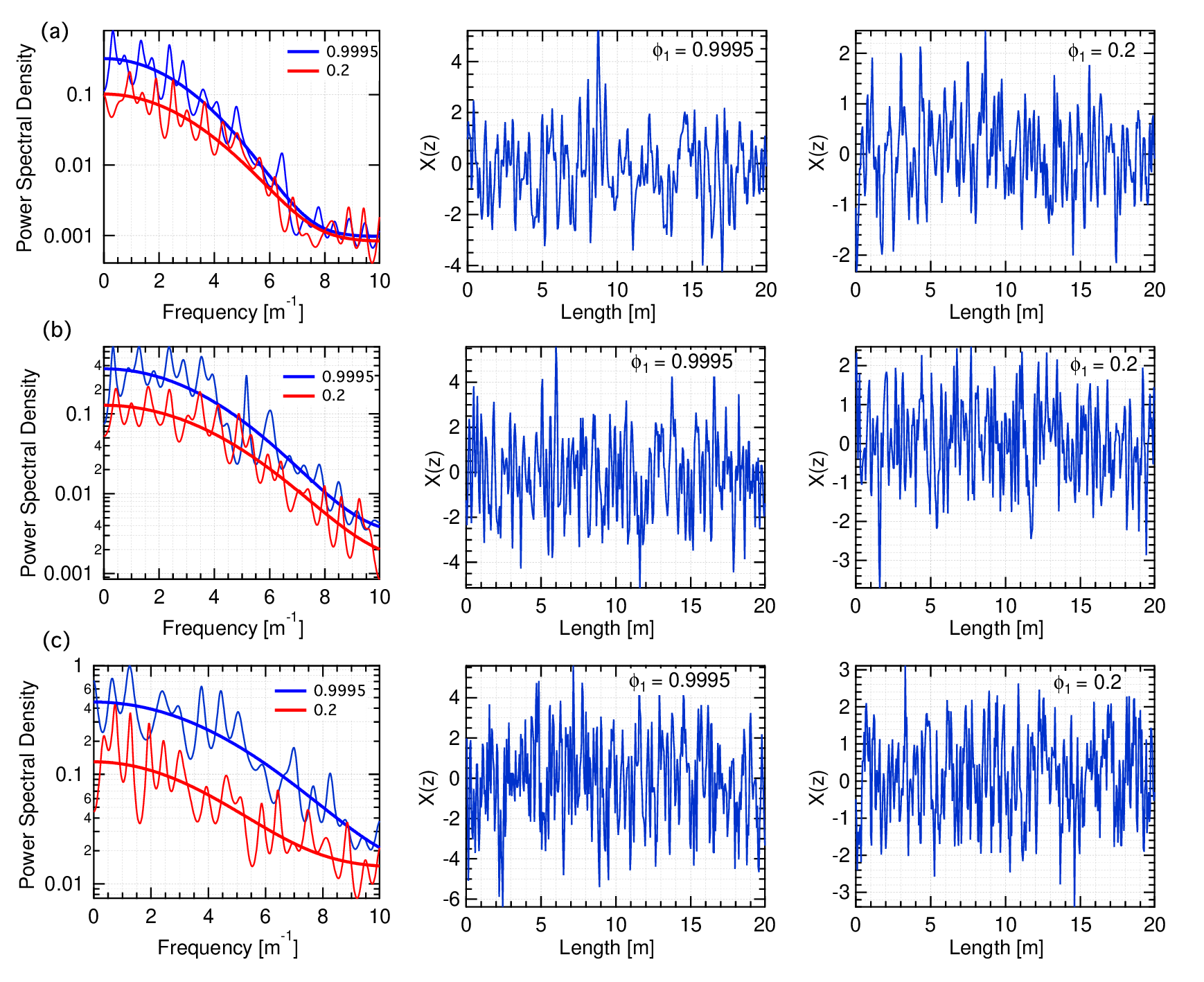}
\caption{Synthetic data test 1. Time series
after diffusion and discrete sampling for $\phi = 0.9995 \, \mathrm{and} \, 0.2$.
The  power spectral densities and the estimated  $\hat{P}_s$ are also presented.
(a) $\sigma = 0.05 \,\mathrm{m}$.
(b) $\sigma = 0.035 \,\mathrm{m}$.
(c) $\sigma = 0.025 \,\mathrm{m}$.}
\label{6_tests}
\end{figure*}

	The test is run using 3 different values for the diffusion length; 0.05, 0.035 and 0.025 m.
	We perform the procedure 100 times for every diffusion length generating a new AR--1 process
	for every repetition.
	This  results in a total of $2\times3\times100=600$ experiments.
	We compare the estimated values
	for $\sigma^2$ with the target values and calculate the mean and RMS value for the estimation.
	In Fig. \ref{6_tests} we show one example for each of the three sets of different
	diffusion length values used presenting the raw time series for both values of $\phi_1$
	and the respective power spectral densities.
	The results for the 6 sets of experiments are presented in Table \ref{table_6_tests1}.

\subsection{Synthetic data test 2}
\begin{figure*}
\vspace*{1mm}
\center
\includegraphics[width = 150mm]{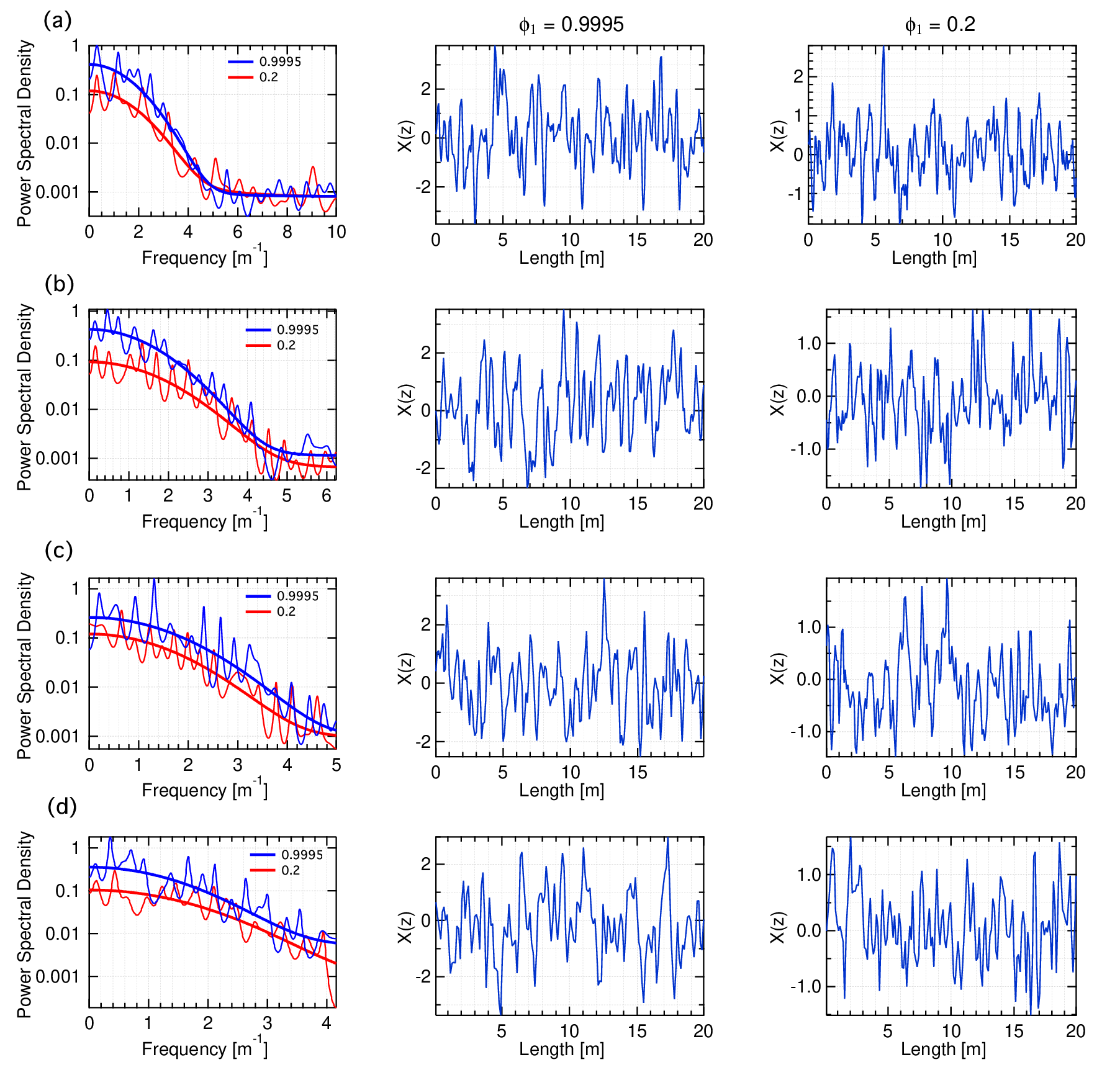}
\caption{Synthetic data test 2. $\sigma = 0.08 \,\mathrm{m}$ for all 4 sets of experiments.
Power spectral densities (column 1) and time series after convolution
with the diffusion filter and discrete sampling  (columns 2 and 3)
for both values of $\phi_1$ are presented.
Here we show the 100th realization of the experiment for every set.
(a) $\Delta = 0.05 \,\mathrm{m}$.
(b) $\Delta = 0.08 \,\mathrm{m}$.
(c) $\Delta  = 0.10 \,\mathrm{m}$.
(d) $\Delta  = 0.12 \,\mathrm{m}$.}
\label{6_tests_sampling}
\end{figure*}

	For the second test we follow a similar procedure as in test 1 generating an AR--1
	process that we then diffuse using a fixed value for the diffusion length. We choose
	$\sigma = 0.08\,\mathrm{m}$.
	The synthetic data sets are then sampled with 4 different resolutions with
	$\Delta = 0.05, 0.08, 0.10 \,\mathrm{and}\,0.12 \,\mathrm{m}$
	The diffusion length is estimated
	and corrected for discrete sampling as described in section 4 of the MT.
	From the spectral estimation point
	of view, the effect of the coarser sampling resolution is equivalent to the ice flow thinning.
	The lower Nyquist frequency caused by the coarser sampling scheme results in an inferior
	estimation of the noise signal (Fig. \ref{6_tests_sampling}).
	The diffusion length estimates that we present in Table \ref{table_6_tests2} indicate that the estimation
	scheme is accurate and insensitive to the memory of the AR--1 process as well as
	the sampling scheme.

	An RMS value of 0.5 cm based on the synthetic data tests is taken into account when
	calculating the confidence intervals in Fig. \ref{mc_tests_results}. The equivalent
	temperature uncertainty of $\pm 0.5\,\mathrm{cm}$ is approximately 1 K for the NorthGRIP site.

%%%%%%%%%%%%%%%%%%%%%%%%%%%%%%%%%%%%%%%%%%%%%%%%%%
%%%%%%%%%%%%%%%%%%%%%%%%%%%%%%%%%%%%%%%%%%%%%%%%%%
%
%\section{Estimation of the uncertainties involved}
%
%	In this section we assess the uncertainties involved in the estimation of the
%	temperature.
%	We identify two main sources of uncertainty. Those related to the parameters
%	of the firn densification model and those related to the ice flow thinning.
%	Each of these two sources of uncertainty is involved at  different stages of the calculation
%	and affect the estimation of the temperature in  different ways.
%
%	Uncertainties related to the firn densification model have an impact on the
%	inferred variability of the temperature signal. Centennial to millennial scale temperature
%	signals as well as the amplitudes of climatic transitions will be affected by this type of uncertainty.
%	On the contrary, the uncertainty associated with
%	the ice thinning correction  affects the slope of the signal.  Relatively small variations
%	in the thinning function can result in differences of several degrees in temperature.
%	However this type of uncertainty can be minimized if estimates of temperature
%	based on other proxies are available and can be used as tie points. In this case
%	the firn diffusion technique can provide combined information on both the temperature history
%	and the ice flow characteristics of the ice core site.

%%%%%%%%%%%%%%%%%%%%%%%%%%%%%%%%%%%%%%%%%%%%%%%%%%
%%%%%%%%%%%%%%%%%%%%%%%%%%%%%%%%%%%%%%%%%%%%%%%%%%
\section{Ice flow thinning effects}
\begin{figure*}
\center
\includegraphics[width = 90mm]{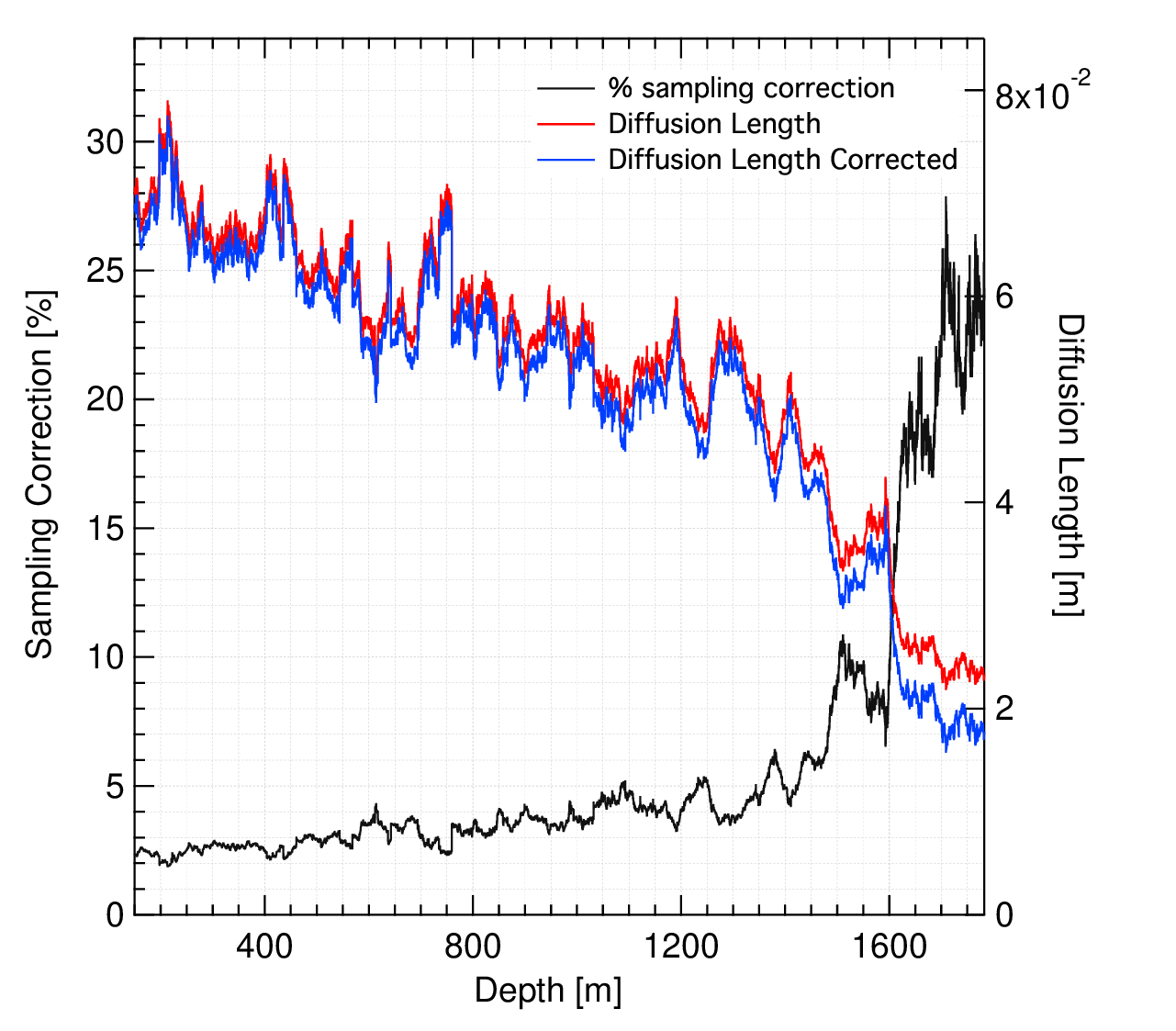}
\caption{Discrete sampling correction}
\label{discrete_sampling01}
\end{figure*}

	The layer thinning induced by the ice flow, impacts the diffusion length estimation mainly in two
	ways. First, due to the discrete sampling scheme as the diffusion length estimation
	moves towards the deeper parts of the core, a single sample averages more years of climate information.
	This effect is essentially taken care of by means of the discrete sampling correction described in
	section 4 of the MT. The term $\sigma_{dis}^2$ is constant with depth. However based on
	Eq. (20) of the MT one can see that the effective correction for discrete sampling scales with
	the total thinning function $S(z)$. In Fig. \ref{discrete_sampling01}, we illustrate
	the effect of this correction with depth.

	Second, the ice flow thinning will result in the diffusion length value decreasing with depth.
	With lower $\sigma^2$ values, a spectrum estimate up to the
	Nyquist frequency $1/2\Delta$ will contain a decreasing
	part of the noise signal  ${\vert \hat{\eta} \left( k \right) \vert} ^{2}$. After a certain depth,
	the sampling resolution is not high enough to resolve ${\vert \hat{\eta} \left( k \right) \vert} ^{2}$.
	The result of this effect is that the estimation of the $P_{\sigma}$ signal requires  an
	assumption about ${\vert \hat{\eta} \left( k \right) \vert} ^{2}$
	and thus it can limit  the  extend to which the diffusion technique can be
	applied to the deeper parts of the core.

\begin{figure*}
\center
\includegraphics[width = 80mm]{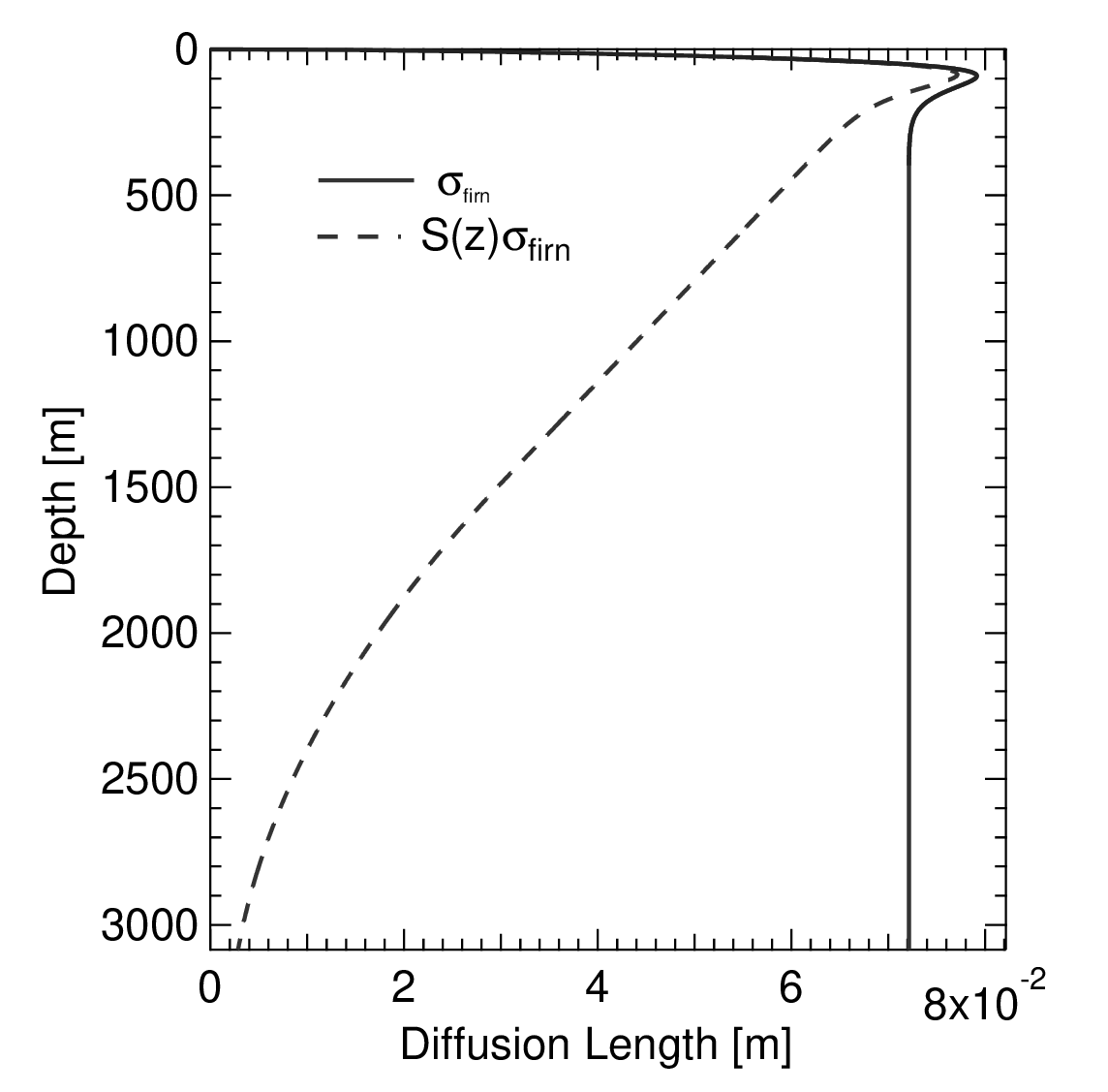}
\caption{Effect of the ice layer thinning on the value of the diffusion length.
	For the calculation of
	$\sigma_{\mathrm{firn}}$ the parameters we used for the H--L model were typical of Holocene
	conditions for the NorthGRIP site:
	$P = 0.7$ Atm, $\rho_0 = 330 \mathrm{\;kgm^{-3}}$,
	$\rho_{\mathrm{CO}} = 804.3 \;\mathrm{kgm^{-3}}$, $T = 242.15$ K, and
	$A = 0.2 \;\mathrm{myr}^{-1}$ ice eq. }
\label{spectral04}
\end{figure*}

\begin{figure*}
\center
\includegraphics[width = 90mm]{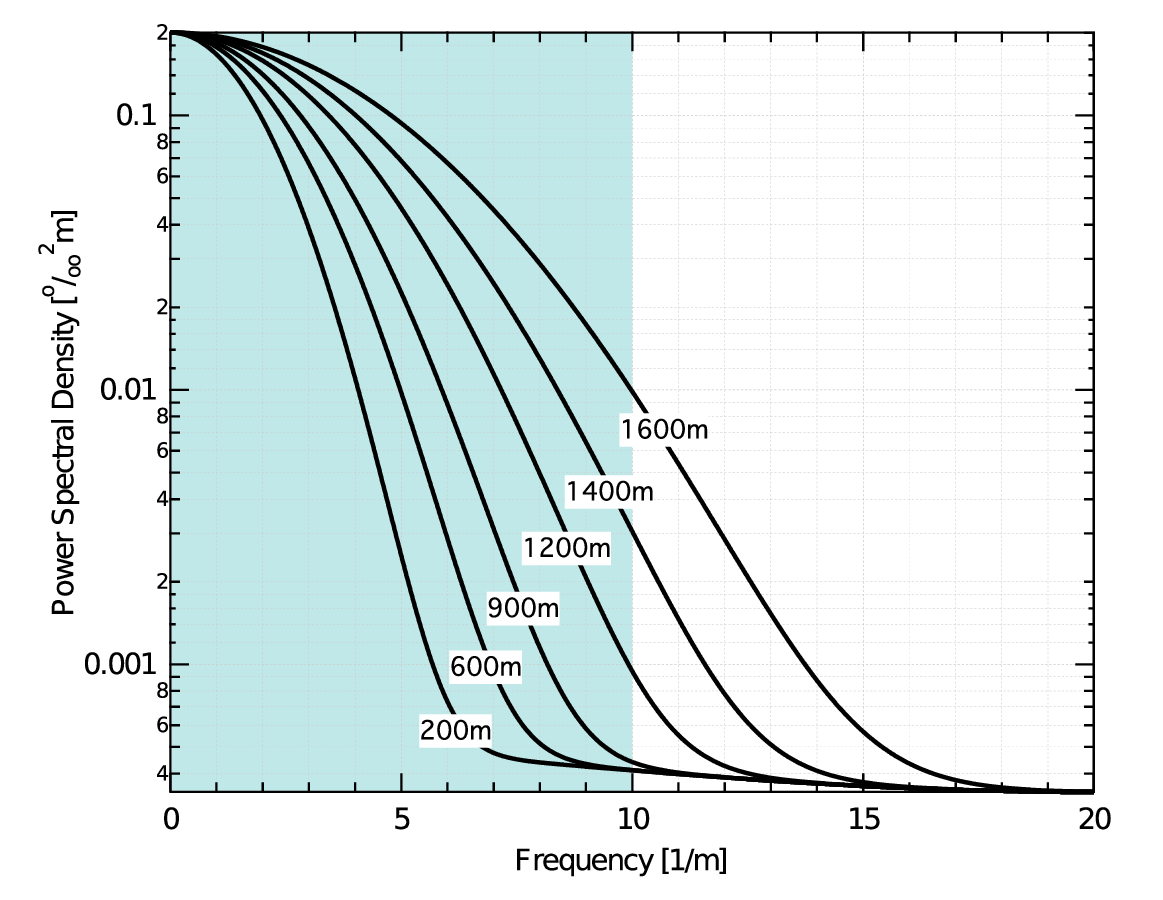}
\caption{Modeled power spectral densities for 6 different depths using diffusion length values from the
calculation of Fig. \ref{spectral04} (Note that all plots represent identical accumulation
rate and temperature forcing at the surface).
We use the spectral model as in Eq. \ref{powersd} with
$q_1 = 0.15, P_{ini} = 0.2$. The blue highlighted area represents the expected spectra for the case of
$\Delta = 5 \mathrm{cm} \, \left(f_{Nyq} = 10 \, \mathrm{m}^{-1}\right)$ while the full range
of the modeled spectra represents the case with
$\Delta = 2.5\, \mathrm{cm} \, \left( f_{Nyq} = 20 \, \mathrm{m}^{-1} \right)$.}
\label{spectral03}
\end{figure*}

	In Fig. \ref{spectral04} we plot the expected diffusion length value assuming
	a simple case of constant temperature
	and accumulation rate at the surface and a certain ice layer thinning history.
	Then, based on this modeled
	diffusion length profile, in Fig. \ref{spectral03}, we calculate six power spectral densities for
	$z = 200, 600, 900, 1200, 1400 \, \mathrm{ and } \, 1600  \, \mathrm{m}$.
	For the spectral calculations
	we use 2 different sampling schemes, $\Delta_1 = 5 \, \mathrm{cm \; and \;} \Delta_2 = 2.5 \,\mathrm{cm}$.
	The plots illustrate the effect of the ice layer thinning as well as the sampling resolution
	on the shape of the power spectral density.
	As depth increases, a progressively smaller portion of the noise signal is resolved.
	Conclusively, for the deeper parts of the core where the ice layer thinning
	has reduced the diffusion length, a higher sampling resolution ($\Delta < 2.5 \, \mathrm{cm}$)
	is preferable  for an even more accurate estimation of $P_s$ and subsequently $\sigma^2$ to be possible.
	For the NorthGRIP reconstruction we present here, we are able resolve the noise signal down to the
	depth of approximately 1450 m.
	For depths higher than 1450 m we make the simplest possible assumption that the noise level
	is equal to the average values we have observed in the Holocene section.

	The accuracy of the ice flow model in inferring the ice thinning function
	has an influence on the uncertainty of our temperature reconstruction.
	The value of the diffusion length of a layer at depth $z$, estimated from the spectral properties of
	a set of \delOx data needs to be corrected for ice flow thinning.
	An inaccurately estimated  thinning function
	affects the inferred values of  $\sigma^2_\text{firn}$ in a linear
	way as we show in Eq. (13) and (20) of the MT. As far as the inferred temperatures
	are concerned,  the ice thinning function impacts the slope of the signal, but has no influence
	on its variability.

\begin{figure*}
\center
\includegraphics[width = 100mm]{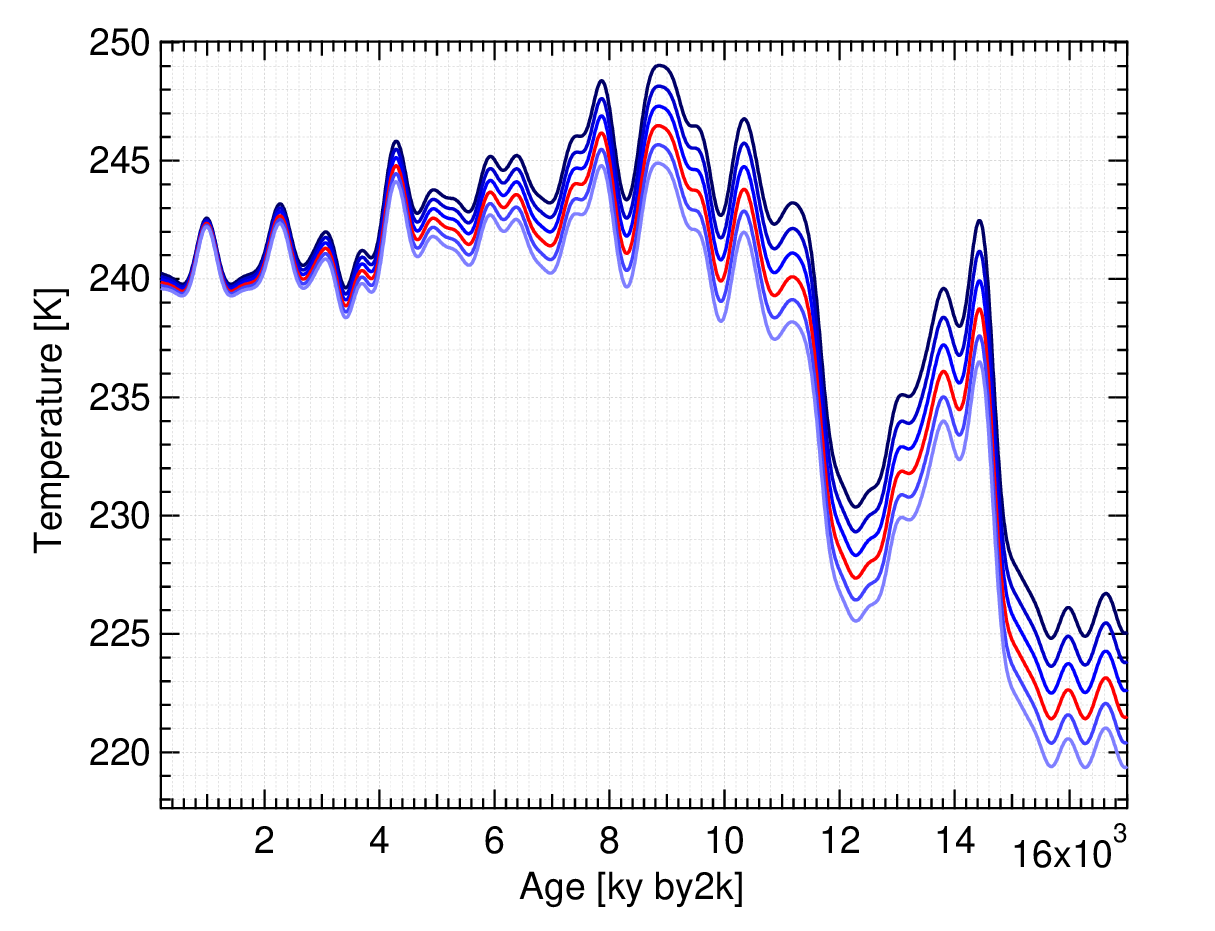}
\caption{Ice thinning uncertainty. Temperature reconstructions
for NorthGRIP using (from top to bottom) $S(2100m) = 0.22, 0.24, 0.26, 0.28, 0.30, 0.32$.
Records filtered with a 500y low-pass filter.}
\label{strain_uncertainty}
\end{figure*}

	In Fig. \ref{strain_uncertainty} we performed the temperature calculation using six different
	scenarios for the ice thinning function $S(z)$. We assume the simple scenario of a
	thinning function that varies linearly
	with depth and a value of $S(z = 2100 \;\text{m})$ equal to  $0.22, 0.24, 0.26, 0.28, 0.30, 0.32$.
	It is apparent that the change in temperature due to thinning function uncertainties can be significant.
	However this type of uncertainty only affects the slope of the temperature signal.
	So,  unrealistic thinning function scenarios are relatively straightforward to rule out.
	Estimates of temperature from other proxies for any point of the record,
	can be useful  in order to select a plausible
	scenario for the thinning function.
	This allows for a more accurate determination of slope of the temperature signal inferred
	by means of the firn diffusion method.

	A direct consequence of this is that the method can potentially be useful in providing combined
	paleotemperature and glaciological information.
	In this study the unrealistically high temperature values we inferred
	for the Holocene climatic optimum pointed to possible
	inaccuracies of the ice thinning function used for the estimation.
	When fixing the temperature gradient
	between the Holocene optimum and present conditions to be approximately 3 K,
	we were able to propose a more likely scenario for the ice thinning function and hence
	the accumulation rate history.
	The temperature reconstruction using the proposed ice thinning
	function for NorthGRIP is presented in red color in Fig. \ref{strain_uncertainty}.

%%%%%%%%%%%%%%%%%%%%%%%%%%%%%%%%%%%%%%%%%%%%%%%%%%
%%%%%%%%%%%%%%%%%%%%%%%%%%%%%%%%%%%%%%%%%%%%%%%%%%
\section{Firn densification uncertainties}
\label{densification_uncertainties}

	Uncertainties related to the densification model affect the estimation of the diffusion
	length  $\sigma^2_{\mathrm{firn}}$. Hereby we examine the
	influence of four parameters involved in the densification--diffusion model. We run
	a set of sensitivity experiments where the four firn densification parameters
	are perturbed in order to create a family of 1000 implementations of the diffusion-densification
	model for each experiment. For all the following sensitivity tests we also consider
	the standard deviation of the diffusion length spectral estimate as calculated in
	sections  \ref{AR_order_selection} and \ref{synthetic_data}.

	The first two parameters we consider are the surface and close--off densities $\rho_0$ and $\rho_{\mathrm{co}}$.
	We perform two sensitivity experiments where  the values of
	$\rho_0$ and $\rho_{\mathrm{co}}$ are drawn from  a Gaussian distribution with a
	defined mean and standard deviation (Table \ref{table_mctests}).
	For the close--off density
	a value of $804 \;  \pm 20 \;\text{kgm}^{-3} \;(1\sigma)$ is used
	\citep{Schwander1988, Jean-Baptiste1998, Johnsen2000}.
	The range of values we choose for $\rho_\text{co}$ brackets within $2\sigma$ the more
	extreme estimates of $775 \; \text{and} \; 840  \;\text{kgm}^{-3}$ shown in
	\cite{Scher1970} and \cite{Stauffer1985} respectively.
	For the surface density we use a  value of
	$320 \pm 40 \;\text{kgm}^{-3} \;(1\sigma)$, based on modern observations of
	the firn column density at NothGRIP. Previous high resolution density observations by
	\cite{Albert2002} for Summit, Greenland indicate that the surface density can
	vary within $\pm 50 \mathrm{kgm}^{-3}$ of its mean value. As a result, with  a $1\sigma$
	of $ 40 \;\mathrm{kgm}^{-3}$  the Gaussian distribution of $\rho_0$
	covers this range adequately in our sensitivity experiments.

\begin{figure*}
\center
\includegraphics[width = 90mm]{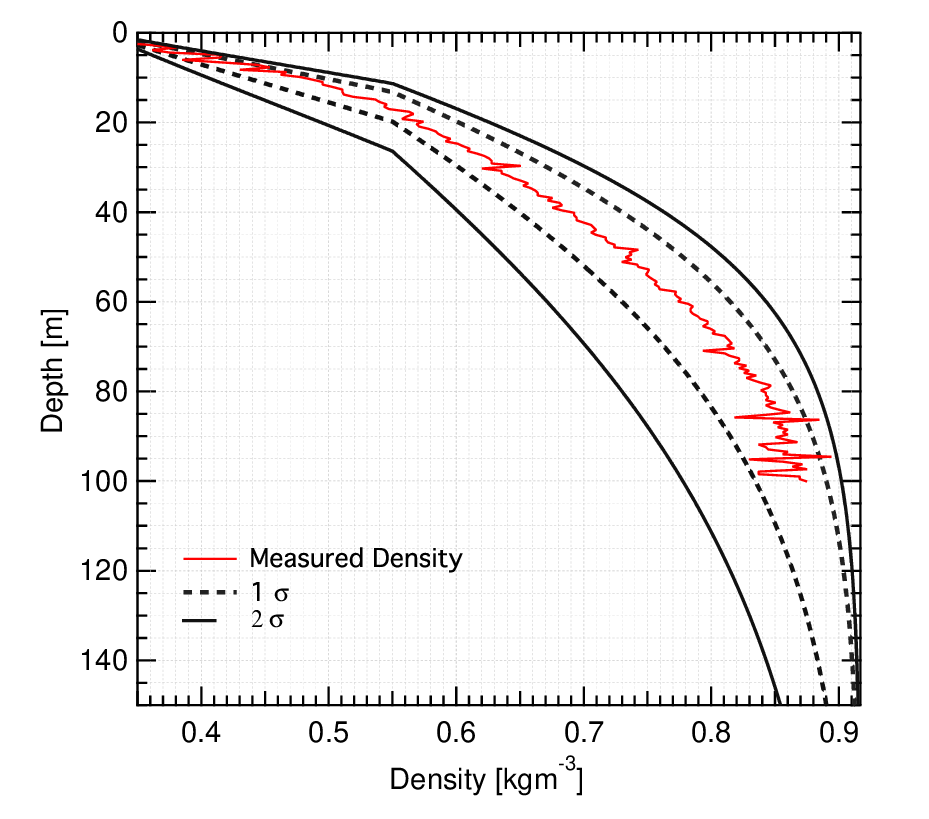}
\caption{Firn density measurements from NorthGRIP (red) compared to implementations of
the H-L densification model with varying values of $f$. Solid and dashed curves represent
$1\sigma$ and $2\sigma$ respectively.}
\label{density_NGRIP_fof1}
\end{figure*}

	Additionally, we include two parameters that describe the dependance
	of the densification rate to temperature. Based on \cite{Herron1980}
\begin{equation}
\frac{\mathrm{d}\rho(z)}{\mathrm{d}t} = K(T)A^{\vartheta} \left( \rho_{\mathrm{ice}} - \rho\left(z\right)\right),
\label{eqhl}
\end{equation}
	where $K(T)$ is a temperature dependent Arrhenius--type densification rate coefficient described by:
\begin{equation}
K(T) = 11 \exp \left( - \frac{10160}{RT} \right)\;\;\; \rho<550 \; \mathrm{kgm}^{-3},
\label{ko_herron}
\end{equation}
and
\begin{equation}
K(T) = 575 \exp \left( - \frac{21400}{RT} \right) \;\;\; \rho\ge 550 \; \mathrm{kgm}^{-3}.
\label{k1_herron}
\end{equation}
	In order to perturb the model we  use the term $K'(t)$ in Eq. (\ref{eqhl}) where
	$K'(T) = fK(T)$ and $f = 1 \pm 0.2 \; (1\sigma)$. This results in a family of density profiles that are used
	for the diffusion length calculation. In Fig. \ref{density_NGRIP_fof1} $1\sigma$
	and $2\sigma$ intervals are illustrated together with firn density measurements from
	NorthGRIP.

\begin{figure*}
\center
\includegraphics[width = 100mm]{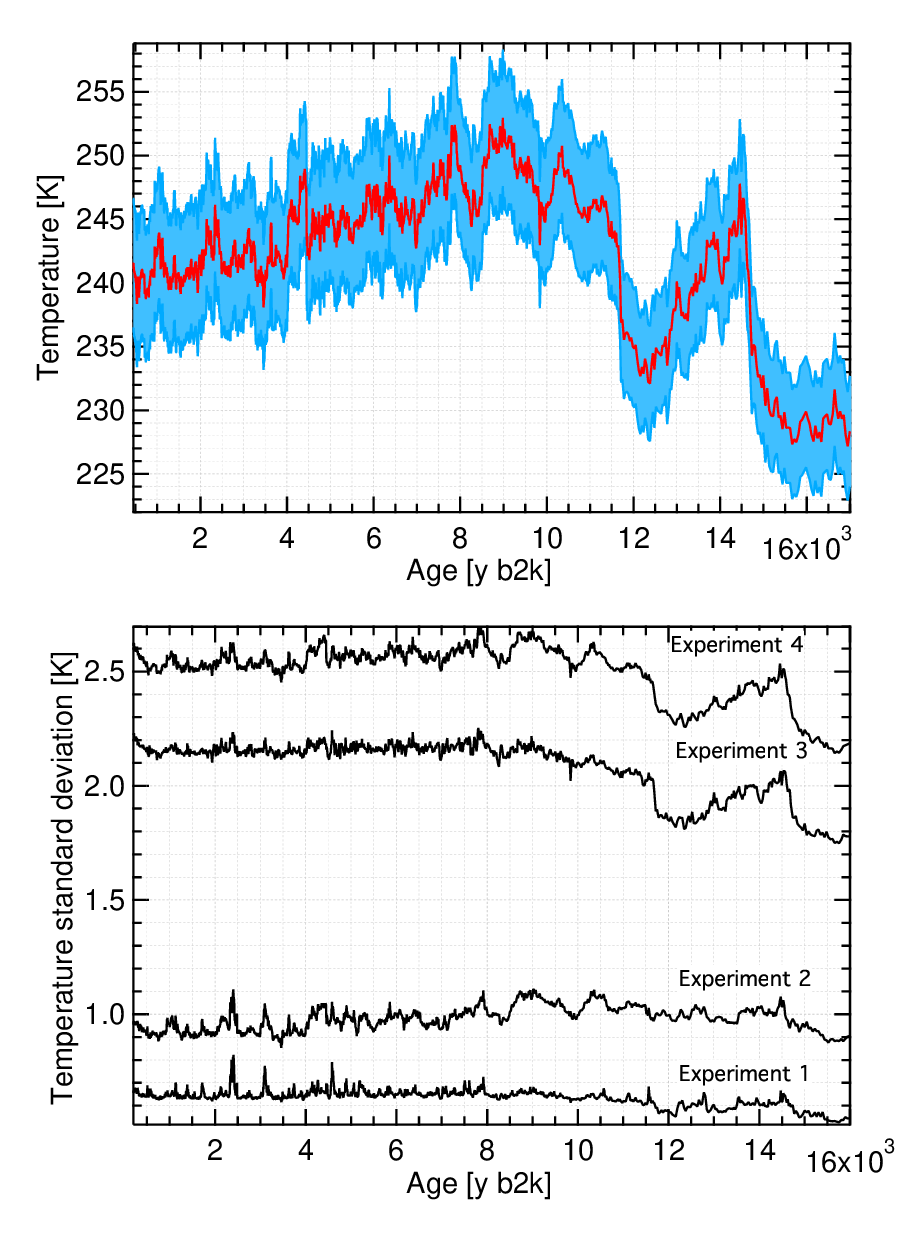}
\caption{Sensitivity tests results. The top panel illustrates the mean temperature history
as calculated from Experiment 4 (all parameters varied) bracketed by the 95\% confidence interval estimated
with the sensitivity test. For the 95\% confidence interval we have also taken into account
the uncertainty of the spectral estimation based on the synthetic data tests and using an
RMS value of $\pm0.5 \,\mathrm{cm}$ that is equivalent to $\approx \, \pm1\,\mathrm{K}$ in temperature.
In the bottom panel we present the value of the standard deviation ($1\sigma$)
for each sensitivity test.}
\label{mc_tests_results}
\end{figure*}

	The results of these sensitivity experiments are illustrated in Fig.\ref{mc_tests_results}.
	Based on these results we conclude that using
	a fixed value for the surface and close--off densities is a plausible approach. The combined uncertainty
	of the $\rho_o$ and $\rho_{\text{co}}$ parameters is in the order of 1 K and thus
	the temperature history we infer is consistent over a wide range of densification parameter
	values. Combining all densification parameters the uncertainty of the estimation is equal to
	$\pm 2.5 \,\mathrm{K}\,(1\sigma)$. Combining this in a Gaussian sense with the spectral estimation uncertainty
	from section \ref{synthetic_data} we get a total of $\pm 2.7 \,\mathrm{K}\,(1\sigma)$.

\bibliographystyle{harvard}
%\bibliographystyle{elsarticle-harv}
%\bibliography{BitexLib.bib}
%\bibliography{/Users/vasilis/Documents/ref's/bib_vasi.bib}
%% else use the following coding to input the bibitems directly in the
%% TeX file.

\begin{center}
\begin{table*}
\centering
\begin{tabular}{r | c c c}
$\sigma \,[\mathrm{cm}]$ &  5  & 3.5 & 2.5   \\ \hline
$\phi_1 = 0.2$ & $4.99 \pm 0.20$ &  $3.47 \pm 0.19$ & $2.30 \pm 0.25$ \\
$\phi_1 = 0.9995$ & $5.06 \pm 0.15$ &  $3.50 \pm 0.13$ & $2.56 \pm 0.39$ \\
\hline
\end{tabular}
\caption{Summary of the synthetic data test 1. The mean and RMS value of the
diffusion length estimation is given in cm for 3 different target values of the diffusion length.
Statistics are  based on 100 realizations for each experiment.}
\label{table_6_tests1}
\end{table*}
\end{center}

\begin{center}
\begin{table*}
\centering
\begin{tabular}{l | c c c c}
 &  \multicolumn{4}{c}{Sampling Interval [cm]}  \\\hline
& 5 & 8 & 10 & 12 \\\hline
$\phi_1 = 0.2$ & $8.1 \pm 0.3$ &  $8.0\pm0.4$ & $8.1\pm0.3$ & $8.2\pm0.7$ \\
 $\phi_1 = 0.9995$ & $8.2\pm0.2$ &  $8.1\pm0.3$ & $8.1\pm0.2$ & $8.1\pm0.5$ \\
\hline
\end{tabular}
\caption{Summary of the synthetic data test 2. The mean and RMS value of the
diffusion length estimation given in cm, is based on 100 realizations of the experiment
for each set of AR-1 coefficient $\phi_1$ and sampling interval $\Delta x$.}
\label{table_6_tests2}
\end{table*}
\end{center}

\begin{table*}
\centering
\begin{tabular}{l | l r r r}
Site & Location & Accum. Rate $[\text{myr}^1]$ & Temperature [C] & $\sigma_{18}^2$ [m]  \\\hline
Dome C & $75^{\circ}06'\text{S}\; 123^{\circ}21'\text{E}$ & 0.027 &  -54.5 & 0.067\\
GISP2 & $72^{\circ}36'\text{N}\; 38^{\circ}30'\text{W}$ & 0.24 & -31.4 & 0.079\\
GRIP & $72^{\circ}35'\text{N}\; 37^{\circ}38'\text{W}$ & 0.23 & -31.7 & 0.0795\\
NEEM & $77^{\circ}45'\text{S}\; 51^{\circ}06'\text{W}$ & 0.2 & -30 & 0.088\\
NorthGRIP & $75^{\circ}10'\text{N}\; 42^{\circ}32'\text{W}$ & 0.207 & -32 & 0.081\\
SipleDome & $81^{\circ}40'\text{S}\; 148^{\circ}46'\text{W}$ & 0.087 & -25 & 0.145\\
South Pole & $90^{\circ}\text{S}\; 00^{\circ}$ & 0.076 & -51 & 0.054\\
Vostoc & $78^{\circ}27'\text{S}\; 10^{\circ}51'\text{E}$ & 0.024 & -55.5 & 0.067\\
\hline
\end{tabular}
\caption{Surface forcing used for the diffusion length calculations in Fig. \ref{diffusionlength_map}. }
\label{sigma_18_core_sites}
\end{table*}

\begin{table*}
\centering
\begin{tabular}{l | c c c r r}
 & $f_o$ & $f_1$ & $\rho_0$ & $\rho_\text{co}$ \\ \hline
Experiment 1 & 1 & 1 & 320 $\text{kgm}^{-3}$ & $804 \pm 20 \text{kgm}^{-3}$ \\
Experiment 2 & 1 & 1 & $320 \pm 40 \; \text{kgm}^{-3}$ & $804 \; \text{kgm}^{-3}$ \\
Experiment 3 & $1 \pm 0.2$ & $1 \pm 0.2$ & 320 \;$\text{kgm}^{-3}$ & $804 \;\text{kgm}^{-3}$ \\
Experiment 4 & $1 \pm 0.2$ & $1 \pm 0.2$ & $320 \pm 40 \; \text{kgm}^{-3}$ & $804 \pm 20 \;\text{kgm}^{-3}$ \\
\hline
\end{tabular}
\caption{Summary of the sensitivity experiments run}
\label{table_mctests}
\end{table*}

\newpage

\end{document}